\documentclass[twocolumn,prl,preprintnumbers,amsmath,amssymb]{revtex4}

\usepackage{graphicx}
\usepackage{dcolumn}
\usepackage{bm}
\usepackage{times}
\usepackage{color}

\begin{document}

\bibliographystyle{unsrt}
\preprint{A. H\"{o}gele {\em et al.}, October 2004}

\title{Voltage-controlled optics of a quantum dot}

\author{Alexander H\"{o}gele}
\affiliation{Center for NanoScience and Department f\"{u}r Physik,
Ludwig-Maximilians-Universit\"{a}t, Geschwister-Scholl-Platz 1,
80539 M\"{u}nchen, Germany}

\author{Stefan Seidl}
\affiliation{Center for NanoScience and Department f\"{u}r Physik,
Ludwig-Maximilians-Universit\"{a}t, Geschwister-Scholl-Platz 1,
80539 M\"{u}nchen, Germany}

\author{Martin Kroner}
\affiliation{Center for NanoScience and Department f\"{u}r Physik,
Ludwig-Maximilians-Universit\"{a}t, Geschwister-Scholl-Platz 1,
80539 M\"{u}nchen, Germany}

\author{Khaled Karrai}
\affiliation{Center for NanoScience and Department f\"{u}r Physik,
Ludwig-Maximilians-Universit\"{a}t, Geschwister-Scholl-Platz 1,
80539 M\"{u}nchen, Germany}

\author{Richard J.\ Warburton}
\affiliation{School of Engineering and Physical Sciences,
Heriot-Watt University, Edinburgh EH14 4AS, UK}

\author{Brian D. Gerardot}
\affiliation{Materials Department, University of California, Santa
Barbara, California 93106, USA}

\author{Pierre M. Petroff}
\affiliation{Materials Department, University of California, Santa
Barbara, California 93106, USA}

\date{\today}



\begin{abstract}
We show how the optical properties of a single semiconductor quantum dot can be controlled with a small dc
voltage applied to a gate electrode. We find that the transmission spectrum of the neutral exciton exhibits two
narrow lines with $\sim 2$~$\mu$eV linewidth. The splitting into two linearly polarized components arises
through an exchange interaction within the exciton. The exchange interaction can be turned off by choosing a
gate voltage where the dot is occupied with an additional electron. Saturation spectroscopy demonstrates that
the neutral exciton behaves as a two-level system. Our experiments show that the remaining problem for
manipulating excitonic quantum states in this system is spectral fluctuation on a $\mu$eV energy scale.
\end{abstract}

\maketitle

In order to realize semiconductor-based schemes for the generation
of single photons \cite{Michler}, the generation of entangled
photons \cite{Benson}, and photonics-based quantum information
processing \cite{Barenco}, it is necessary to generate and
manipulate optically active quantum states. Although atoms, ions
and molecules are possible hosts for these processes, quantum dots
have the advantage that the solid state matrix acts as a built-in
trap. In addition, semiconductor heterostructure technology allows
considerable control over the continuum states at energies above
the quantum dot energy levels. We demonstrate here how this
capability can be exploited to control the polarization, the
oscillator strength and the energy of the optical transitions in a
single quantum dot simply with a dc voltage. In particular, we are
able to turn an exchange energy on and off. This level of
tunability has never been achieved previously on any quantum
object, atom, molecule or quantum dot. Furthermore, our
experiments provide compelling evidence that the ground state
exciton behaves as a highly coherent two-level system.

Significant progress in quantum dot optics has been made with natural quantum dots, localizing fluctuations in a
disordered quantum well \cite{Gammon}, leading to the recent demonstration of an optical CROT logic gate
\cite{Li}. For our own experiments, we employ a self-assembled quantum dot which has a much stronger confining
potential than a natural dot, leading to a significantly longer excitonic coherence time of several hundred ps
\cite{Borri,Bayer}. The oscillator strength of a self-assembled quantum dot is about 10, smaller than that of a natural quantum dot \cite{Guest}, making the controlled creation and detection of excitonic quantum states very
challenging. We report here the first detection of ground state exciton creation in a single self-assembled
quantum dot through a reduction in the transmission coefficient. This is a very direct experimental technique as
it does not involve a third state as is inevitably the case in a luminescence-based method. Furthermore, it
profits from all the advantages of laser spectroscopy, namely high spectral resolution, precise line shape
determination and direct access to the oscillator strength.

The InGaAs quantum dots are embedded in a GaAs matrix and emit around 1.3~eV. The dots are
positioned 25~nm above a metallic-like layer (n$^{+}$-GaAs) and 150~nm beneath a Schottky barrier on the surface
allowing us to modulate the properties with the gate voltage, the voltage applied to the Schottky
contact. The gate voltage induces a vertical electric field which shifts the exciton energy through the
Stark effect \cite{Alen}. By changing the potential of the quantum dot relative to the back contact, we can control the occupation. There is a pronounced Coulomb blockade both in the
capacitance \cite{Miller} and in the photoluminescence \cite{Nature}, allowing us to control unambiguously the
electron number and excitonic charge, respectively. Optical absorption was measured with a narrow band (5~MHz)
laser, detecting the transmission with a Ge p-i-n photodiode placed directly beneath the sample at 4.2~K, as
shown in Fig.~1(a). The Gaussian beam from the laser was focussed to a 1.2~$\mu$m full-width-at-half-maximum
(FWHM) spot. The dot density in our sample is $\sim$~5$\cdot$10$^{9}$~cm$^{-2}$, implying that several tens of
dots lie in the focus. However, owing to the narrow laser line, quantum dots can be addressed individually. With
the given spot size, a combination of the large sample refractive index and the homogeneous broadening implies
that the change in transmission at resonance is small, and therefore, as in single molecule \cite{Moerner,Kador}
and single ion \cite{Wineland} transmission experiments, a differential technique was adopted. In addition to a
dc voltage, a square wave voltage with peak-to-peak amplitude 100~mV was applied to the gate. Spectroscopy was
performed by sweeping the dc gate voltage at constant laser energy measuring the differential transmission with
a lock-in amplifier. Initially, the power was kept below 10~nW to avoid saturation effects. Fig.~1(b)
shows a schematic representation of the quantum mechanical states involved in the creation of the neutral
exciton X${^0}$ and the singly charged exciton X$^{1-}$, and Fig.~1(c) denotes the levels involved in the
optical transitions.

\begin{figure}[t]
\begin{center}
\includegraphics[scale=0.8]{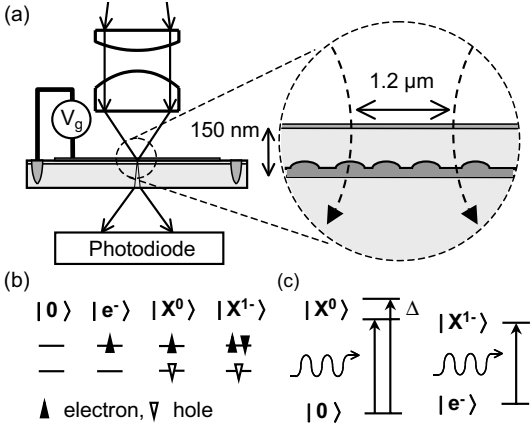}
\end{center}
\caption{\label{fig1}(a) The optical microscope (to scale). Laser
light is delivered with an optical fibre (not shown); the light is
collimated and then focussed with an aspherical lens with
numerical aperture 0.55 onto the sample. The FWHM spot size was
measured to be 1.2~$\mu$m. Transmitted light is detected with an
in situ Ge p-i-n photodiode. (b) Quantum mechanical states in a
single quantum dot: $| 0 \rangle$ is the vacuum state, $| {\rm
e}^{-}\rangle$ the single electron state, $| {\rm X}^{0}\rangle$
the neutral exciton state, and $| {\rm X}^{1-}\rangle$ the singly
charged exciton state. (c) The level diagrams for the creation of
a neutral exciton and a singly charged exciton. The neutral
exciton is split by $\Delta$ through the anisotropic electron-hole
exchange interaction.}
\end{figure}

\begin{figure}[t]
\begin{center}
\includegraphics[scale=0.9]{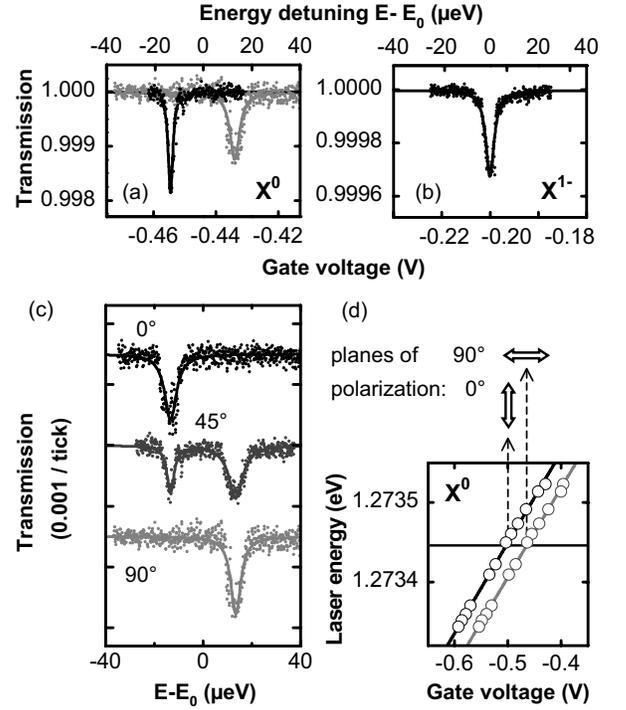}
\end{center}
\caption{\label{fig2}(a) Differential transmission of the neutral
exciton in a single self-assembled quantum dot$^{*}$. The two
curves were recorded with orthogonal linear polarization. (b)
Differential transmission for the singly charged exciton,
X$^{1-}$. (c) Differential transmission for X$^{0}$ for three
directions of the linear polarization. (d) Voltage-induced
Stark-shift of the two resonance energies of the neutral exciton.
At a fixed photon energy, the polarization can be switched between
the two orthogonal linear orientations through the gate voltage.
The detuning in (a), (b) and (c) was achieved at constant laser
wavelength by sweeping the gate voltage. The sample was at 4.2~K.
$^{*}$Note: Experimental data in Fig.s~2(a), (b) and (c) were
recorded on one quantum dot; those in Fig.~2 (d), and in Fig.s~3
and 4 were recorded on another quantum dot.}
\end{figure}

Fig.~2(a) shows differential transmission spectra of a single
quantum dot for voltages at which the photoluminescence exhibits
only recombination of the neutral exciton. The gate voltage is
converted into a detuning, $E-E_0$, by repeating the measurements
at several known laser wavelengths. In a narrow range of voltage,
the exciton energy varies linearly with gate voltage (Fig.~2(d)).
The neutral exciton exhibits two pronounced Lorentzian-shaped
resonances, separated by $\Delta=$ 27~$\mu$eV, as shown in
Fig.~2(a) and (c). We find values of $\Delta$ ranging from
10~$\mu$eV to 42~$\mu$eV for the ten different quantum dots
measured in this sample. As shown in Fig.~2(c), the two neutral
exciton lines are linearly polarized in orthogonal directions. The
fine structure splitting $\Delta$ arises through the electron-hole
exchange interaction \cite{Bayereh}, its magnitude indicating that
a shape anisotropy makes the dominant contribution \cite{Bester}.
The main point is that the quantum dot exciton has a natural
linear polarization basis. Our results show that we can exploit
this feature to prepare very precisely the excitonic state of the
quantum dot.

By controllably addressing the $x$- and $y$-polarized excitons, the dot can emit only one particular
polarization. This is a very attractive concept for a single photon source as it allows control over the
polarization of the emitted photon. However, without the electric field, the two polarizations have different
energies such that there is an unwanted correlation between the polarization and the photon energy. With our
field-effect device this correlation can be removed. Fig.~2(d) demonstrates how the $x$- and $y$-polarized
excitons can have exactly the same energy by applying two slightly different gate voltages, an ideal feature for
a single photon source.

It would clearly be desirable to change also the basis of the quantum dot exciton to generate not
just linearly polarized photons but also circularly polarized photons. Our structure allows us to do exactly
this. By making the gate voltage slightly more positive, the electronic Coulomb blockade allows us to populate
the quantum dot with a single electron in the absence of an optical excitation. In this case, as shown in
Fig.~2(b), the doublet in the transmission spectrum is replaced by a single line. The reason for this is that
the exciton generated by photon absorption now has a filled $s$ shell (Fig.~1(b)) and therefore total electron
spin zero such that the exchange interaction with the spin-$\frac{3}{2}$ hole vanishes . While this argument has
been invoked to explain several experiments \cite{Bayereh,Finley, Tischler}, our own experiment provides the first proof that the neutral exciton's fine structure disappears
on charging. It is now possible to control the polarization selection rules of the exciton by controlling the
quantum state of the resident electron. Equivalently, a measurement of the spin state of the resident electron
can be made with photon absorption. We demonstrate this concept by applying a magnetic field which splits the
spin-up and spin-down electron states through the Zeeman energy. The exciton splits into two components with
right- and left-handed circular polarizations, exactly as expected in the absence of an exchange interaction, as
shown in Fig.~3(a). Again, by applying slightly different gate voltages, the right- and left-circular polarized
excitons at a fixed magnetic field can have exactly the same energy. Alternatively, the two states can be tuned
into resonance at constant voltage and laser energy with a magnetic field, as demonstrated in Fig.~3(b). Our
experiment demonstrates therefore that charging with a single electron has a profound effect on the excitonic
states, causing the features related to exchange to disappear. Furthermore, the absorption can be turned
off completely by applying a voltage such that the dot contains two electrons \cite{Richard}. In this case, the
ground state is occupied with two electrons and further occupation is forbidden through the Pauli principle.

\begin{figure}[b]
\begin{center}
\includegraphics[scale=0.9]{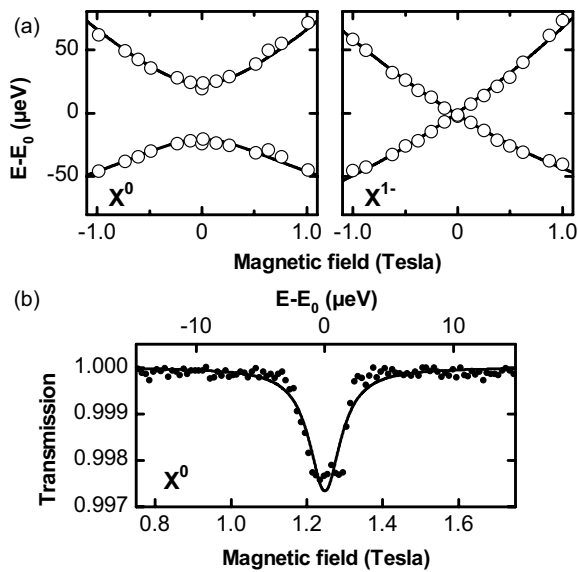}
\end{center}
\caption{\label{fig3}(a) Magnetic dispersion of the neutral
exciton, X$^0$ (left) and the singly-charged exciton, X$^{1-}$
(right). The solid lines are fits to the data with detuning
$\Delta E=\pm \frac{1}{2} g^* \mu_B B+\beta B^2$ for the X$^{1-}$,
and $\Delta E=\pm \frac{1}{2} \sqrt{\Delta^2+(g^* \mu_B
B)^2}+\beta B^2$ for the X$^{0}$ taking the exciton $g$-factor
$g^*=1.8$, the anisotropic electron hole energy
$\Delta=42$~$\mu$eV, and the diamagnetic shift
$\beta=8.7$~$\mu$eV/T$^2$. (b) Differential transmission versus
magnetic field. The resonance corresponds to one branch of the
spin-split neutral exciton.}
\end{figure}

For the manipulation of quantum states, it is desirable to work
with a highly coherent two-level system. Up until now, the
two-level nature of a self-assembled quantum dot exciton has been
probed by detecting Rabi oscillations either on the excited
exciton, relaxation to the ground state exciton providing a
convenient detection scheme \cite{Kamada}, or on the ground state
exciton in a vertical electric field, relying on electron and hole
tunneling to generate a photocurrent \cite{Zrenner}. In both
cases, the detection process compromises the coherence, a problem
we can avoid with saturation spectroscopy. The experiment involves
measuring the maximum contrast in the transmission resonance as a
function of laser power. Saturation corresponds to a reduction in
the contrast. The results, plotted in Fig.~4(a), show how a laser
power of just 100~nW is sufficient to saturate the neutral exciton
transition. We consider two different interpretations, one
involving a coherent two-level system, the other a shelving state.
In the first scenario, a photon excites an exciton which remains
in resonance with the laser such that a subsequent photon can
stimulate recombination. In this case, saturation arises when
stimulated emission dominates over spontaneous emission because in
this limit the detector signal is unchanged in the presence of
resonant absorption. In the second scenario, an exciton relaxes
quickly after its creation into an intermediate shelving state
where it is no longer resonant with the laser. In this case,
saturation arises because the exciton has a finite lifetime. At
high power the system spends a large fraction of the time in the
shelving state where there is zero chance of photon absorption.
The crucial differences between the two models are that saturation
occurs at roughly a factor of three higher power in the shelving
model than in the two-level model, and that the functional
dependencies of the absorption on power are completely different.
Applying the density-matrix formalism to a two-level system, we
find the elegant result for the maximum absorption $\alpha(P)$ as
a function of laser power $P$:
$\alpha(P)/\alpha(0)=1/[1+2\alpha(0)\tilde{N} \tau/\hbar]$ where
$\tilde{N}$ is the photon flux through the quantum dot plane, and
$\tau$ is the radiative lifetime. We have determined $\tau=800 \pm
100$~ps both from the lower power limit of the transmission
spectroscopy and from direct measurements of the radiative decay
allowing us to predict the behavior at high power. As shown in
Fig.~4(a), this two-level analysis gives an excellent fit to the
experimental data. Conversely, in the shelving model,
$\alpha(P)/\alpha(0)=1-\exp[-\hbar/(\alpha(0)\tilde{N}\tau)]$.
With $\tau=800$~ps, the shelving model gives a very poor account
of the saturation characteristics (Fig.~4(a)). A fit can be
obtained in the shelving model only by increasing the radiative
lifetime to 2.4~ns, an unreasonably long lifetime, and even in
this case, an exponential decay gives a much poorer fit to the
data than the two-level model. Remarkably therefore, the
self-assembled quantum dot displays the saturation characteristics
of a two-level system.

The FWHMs of the neutral excitons are 2.3~$\mu$eV and 5.0~$\mu$eV
in Fig.~2. Other dots give similar results, although it is not
always the case that one resonance is broader than the other. The
smallest linewidth we have recorded so far is 1.6~$\mu$eV. In the
ultimate limit of lifetime broadening, the coherence time is equal
to the radiative lifetime, corresponding to a FWHM of 0.8~$\mu$eV
for our dots. Hence, although our resonances are very sharp, we
have not yet achieved the lifetime broadening limit. In order to
probe the broadening mechanism we measured the absorption as a
function of time for constant laser energy, starting with the
quantum dot in exact resonance with the laser. The results are
plotted in Fig.~4(b) which shows also the transmission signal
measured when the laser energy is several linewidths away from the
exciton resonance. On resonance, the absorption signal has quite
large fluctuations whereas off resonance there is just small
random noise. The time constant for this particular experiment was
1 s. The experiment demonstrates that there are fluctuations in
the absorption spectrum of the quantum dot on this time scale. The
results are entirely consistent with a temporal fluctuation of the
resonance position. The absorption signal at constant laser energy
fluctuates as the dot moves in and out of resonance. The
statistical nature of the fluctuations is shown in the histogram
of absorption amplitude, Fig.~4(c). Assuming that on a much
shorter time scale, the resonance is lifetime broadened with a
width of 0.8 $\mu$eV, the data in Fig.~4(c) allow us to deduce
that spectral fluctuation contributes approximately 0.5 $\mu$eV to
the resonance linewidth. Integrating for longer times, as was the
case in the data of Fig.s~2 and 3 (time constant 50 s in
Fig.s~2(a), 2(c) and 2 s in Fig.s~2(b), 3(b)), the resonance is
broadened further, between about 1 and 3 $\mu$eV, by the spectral
fluctuation. On the time scale of the experiment the back contact
is unlikely to be the origin of the spectral fluctuation given its
high conductivity. Instead, preliminary results suggest that the
spectral fluctuation is less in a magnetic field, implying perhaps
that an interaction between the electronic spin and the nuclear
spins in the quantum dot is important.

\begin{figure}[t]
\begin{center}
\includegraphics[scale=0.9]{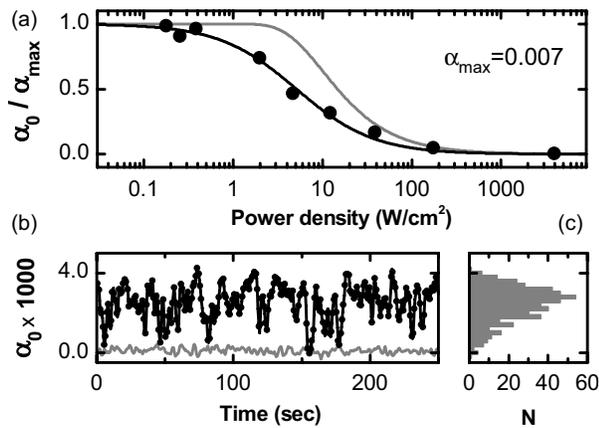}
\end{center}
\caption{\label{fig4}(a) The differential absorption at resonance
plotted against laser power. The solid black line is the result
expected for a two-level system taking a radiative lifetime of
800~ps. The solid grey line models the power dependence of an
absorbing system including an intermediate shelving state, again
taking  a radiative lifetime of 800~ps. (b) The differential
absorption signal as a function of time at zero detuning (black)
and several linewidths away from the exciton resonance (grey). (c)
Histogram of the differential absorption amplitude signal at zero
detuning.}
\end{figure}

In summary, we have demonstrated that a self-assembled quantum dot
can be used to prepare high-fidelity excitonic states with an
unprecedented degree of tunability. The tuning is achieved simply
through a voltage. This property was demonstrated by applying high
resolution laser spectroscopy to a single quantum dot. Our results
demonstrate that the remaining factor limiting the exploitation of
these results for the controlled manipulation of quantum states is
spectral fluctuation on a $\mu$eV energy scale.

We would like to thank J\"{o}rg P. Kotthaus, Harald Weinfurter and
Ian Galbraith for helpful discussions. Financial support for this
work was provided in Germany by DFG grant no.\ SFB~631 and in the
UK by the EPSRC and The Royal Society.


\begin{thebibliography}{}

\bibitem{Michler} P. Michler {\em et al.}, Science {\bf 290}, 2282 (2000); E. Moreau {\em et al.}, Phys. Rev. Lett. {\bf 87}, 1836011 (2001); Z. Yuan {\em et al.}, Science {\bf 295}, 102 (2001); C. Santori {\em et al.}, Nature (London) {\bf 419}, 594 (2002).

\bibitem{Benson} O. Benson {\em et al.}, Phys. Rev. Lett. {\bf 84}, 2513 (2000).

\bibitem{Barenco} A. Barenco {\em et al.}, Phys. Rev. Lett. {\bf 74}, 4083 (1995); F. Troiani, U. Hohenester, and E. Molinari, Phys. Rev. B {\bf 62}, R2263 (2000); E. Biolatti {\em et al.}, Phys. Rev. Lett. {\bf 85}, 5647 (2000).

\bibitem{Gammon} D. Gammon {\em et al.}, Phys. Rev. Lett. {\bf 76}, 3005 (1996).

\bibitem{Li} X. Li {\em et al.}, Science {\bf 301}, 809
(2003).

\bibitem{Borri} P. Borri {\em et al.}, Phys. Rev. Lett. {\bf 87}, 157401 (2001).

\bibitem{Bayer} M. Bayer and A. Forchel, Phys. Rev. B {\bf 65}, 041308(R) (2002).

\bibitem{Guest} J. R. Guest {\em et al.}, Phys. Rev. B {\bf 65}, 241310(R) (2002).

\bibitem{Alen} B. Al\'{e}n {\em et al.}, Appl. Phys. Lett. {\bf 83}, 2235 (2003).

\bibitem{Miller} B. T. Miller {\em et al.}, Phys.
Rev. B {\bf 56}, 6764 (1997).

\bibitem{Nature} R. J. Warburton {\em et al.}, Nature (London) {\bf 405},
926 (2000).

\bibitem{Moerner} W. E. Moerner and L. Kador, Phys. Rev. Lett. {\bf 62}, 2535
(1989).

\bibitem{Kador} L. Kador {\em et al.}, J. Chem.
Phys. {\bf 111}, 8755 (1999).

\bibitem{Wineland} D. J. Wineland, W. M. Itano, and J. C. Bergquist, Opt. Lett. {\bf 12}, 389 (1987).

\bibitem{Bayereh} M. Bayer {\em et al.}, Phys. Rev. B {\bf 65}, 195315 (2002).

\bibitem{Bester} G. Bester, S. Nair, and A. Zunger, Phys. Rev. B {\bf 67}, 161306(R)
(2003).

\bibitem{Finley} J. J. Finley {\em et al.}, Phys. Rev. B {\bf 66},
153316 (2002).

\bibitem{Tischler} J. G. Tischler {\em et al.}, Phys. Rev.
B {\bf 66}, 081310(R) (2002).

\bibitem{Richard} R. J. Warburton {\em et al.}, Phys. Rev. Lett. {\bf 79},
5282 (1997).

\bibitem{Kamada} H. Kamada {\em et al.}, Phys.
Rev. Lett. {\bf 87}, 246401 (2001).

\bibitem{Zrenner} A. Zrenner {\em et al.}, Nature (London) {\bf 418}, 612 (2002).

\end{thebibliography}
\end{document}